\documentclass[sigconf]{aamas}

\usepackage{balance} 
\usepackage{arydshln}
\newtheorem{problem}[]{Problem}
\newtheorem{prompt}[]{Definition Prompt}

\makeatletter
\gdef\@copyrightpermission{
  \begin{minipage}{0.2\columnwidth}
   \href{https://creativecommons.org/licenses/by/4.0/}{\includegraphics[width=0.90\textwidth]{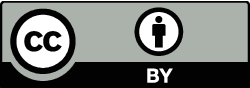}}
  \end{minipage}\hfill
  \begin{minipage}{0.8\columnwidth}
   \href{https://creativecommons.org/licenses/by/4.0/}{This work is licensed under a Creative Commons Attribution International 4.0 License.}
  \end{minipage}
  \vspace{5pt}
}
\makeatother

\setcopyright{ifaamas}
\acmConference[AAMAS '25]{Proc.\@ of the 24th International Conference
on Autonomous Agents and Multiagent Systems (AAMAS 2025)}{May 19 -- 23, 2025}
{Detroit, Michigan, USA}{Y.~Vorobeychik, S.~Das, A.~Nowé  (eds.)}
\copyrightyear{2025}
\acmYear{2025}
\acmDOI{}
\acmPrice{}
\acmISBN{}

\definecolor{mycitecolor}{RGB}{71, 191, 38}
\definecolor{mylinkcolor}{RGB}{40, 115, 201}

\acmSubmissionID{1140}

\title[AAMAS-2025]{Leveraging Large Language Models for Effective and Explainable Multi-Agent Credit Assignment}

\author{Kartik Nagpal}
\affiliation{
  \institution{University of California Berkeley}
  \city{Berkeley, CA}
  \country{United States of America}}
\email{kartiknagpal@berkeley.edu}

\author{Dayi Dong}
\affiliation{
  \institution{University of California Berkeley}
  \city{Berkeley, CA}
  \country{United States of America}}
\email{dayi.dong@berkeley.edu}

\author{Jean-Baptiste Bouvier}
\affiliation{
  \institution{University of California Berkeley}
  \city{Berkeley, CA}
  \country{United States of America}}
\email{bouvier3@berkeley.edu}

\author{Negar Mehr}
\affiliation{
  \institution{University of California Berkeley}
  \city{Berkeley, CA}
  \country{United States of America}}
\email{negar@berkeley.edu}

\begin{abstract}
Recent work, spanning from autonomous vehicle coordination to in-space assembly, has shown the importance of learning collaborative behavior for enabling robots to achieve shared goals. A common approach for learning this cooperative behavior is to utilize the centralized-training decentralized-execution paradigm. However, this approach also introduces a new challenge: how do we evaluate the contributions of each agent's actions to the overall success or failure of the team. This ``credit assignment'' problem has remained open, and has been extensively studied in the Multi-Agent Reinforcement Learning~(MARL) literature. In fact, humans manually inspecting agent behavior often generate better credit evaluations than existing methods. We combine this observation with recent works which show Large Language Models~(LLMs) demonstrate human-level performance at many pattern recognition tasks. Our key idea is to reformulate credit assignment to the two pattern recognition problems of sequence improvement and attribution, which motivates our novel Large Language Model Multi-agent Credit Assignment~(LLM-MCA) method. Our approach utilizes a centralized LLM reward-critic which numerically decomposes the environment reward based on the individualized contribution of each agent in the scenario. We then update the agents' policy networks based on this feedback. We also propose an extension LLM-TACA where our LLM critic performs explicit task assignment by passing an intermediary goal directly to each agent policy in the scenario. Both our methods far outperform the state-of-the-art on a variety of benchmarks, including Level-Based Foraging, Robotic Warehouse, and our new ``Spaceworld'' benchmark which incorporates collision-related safety constraints. As an artifact of our methods, we generate large trajectory datasets with each timestep annotated with per-agent reward information, as sampled from our LLM critics. By making this dataset available, we aim to enable future works which can directly train a set of collaborative, decentralized policies offline.\footnote{Project Page is available at \url{http://iconlab.negarmehr.com/LLM-MCA/}}
\end{abstract}

\keywords{Credit Assignment, Task Allocation, Multi-Agent Reinforcement Learning, Large Language Models, Foundation Models}

\newcommand{\BibTeX}{\rm B\kern-.05em{\sc i\kern-.025em b}\kern-.08em\TeX}

\newcommand{\nb}{\textsuperscript{*}}

\begin{document}

\pagestyle{fancy}
\fancyhead{}

\maketitle 

\begin{figure}[H]
    \centering
    \includegraphics[width=\linewidth]{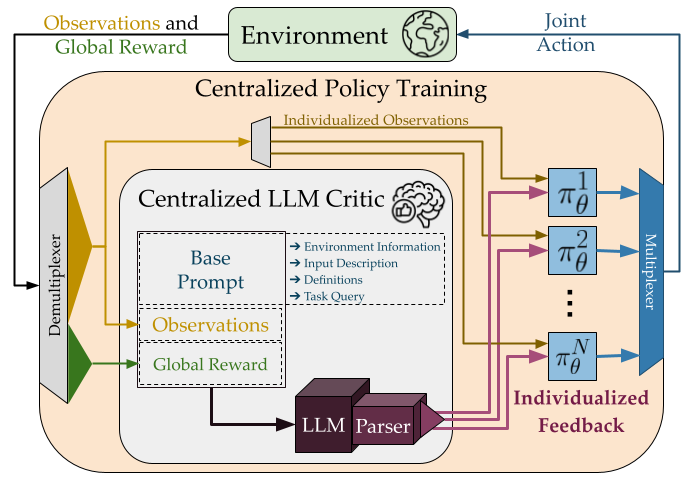}
    \caption{Overall architecture diagram for our LLM-MCA and LLM-TACA methods. \textmd{Our centralized training architecture utilizes a centralized LLM-critic instantiated with our base prompt~(environment description, our definitions, and task query). At each timestep, we update our LLM-critic with the global reward and latest observations from the environment. We then update our agents' policies with the individualized feedback from our critic.}}
    \label{fig:diagram}
    \Description{diagram}
\end{figure}

\section{Introduction}

\begin{figure*}[t!]
    \centering
    \includegraphics[width=0.9\linewidth]{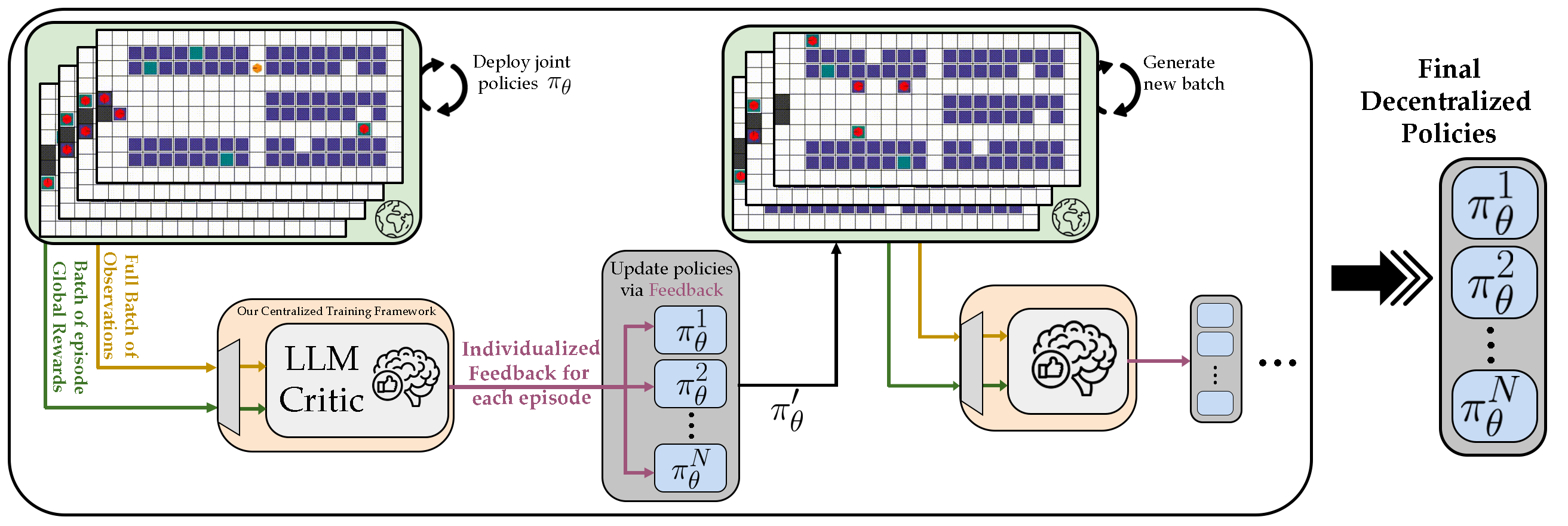}
    \caption{Diagram for our batch-training process with our LLM-MCA method. \textmd{Our centralized training process allows us to provide entire batches of trajectories at once to our centralized LLM-critic. Our LLM-MCA critic then generates individualized feedback for each agent, which we use to update their policies. After training, we no longer need our LLM-critic, and directly deploy our trained, decentralized agent policies.}}
    \label{fig:process}
    \Description{diagram}
\end{figure*}

Many real-world scenarios can be posed as multi-agent cooperative problems, where multiple agents are either required to collaborate or benefit from collaborating towards a shared goal. For instance, moving a heavy object which requires two robots to lift simultaneously, or minimizing city traffic which benefits from vehicles coordinating their routes. While game-theoretic approaches have showcased remarkable results for some multi-agent problems~\cite{bhatt2023efficient}, they rarely focus on collaborative settings. This has prompted many researchers to use Multi-Agent Reinforcement Learning~(MARL)~\cite{weiss1995distributed, zhang2021multi} to coordinate large numbers of agents in complex scenarios. 

The Centralized-Training Decentralized-Execution~(CTDE) paradigm is a popular framework in MARL where agents are trained together via a centralized critic with access to the global states, actions, and rewards~\cite{xu2022maddpg, foerster2016learning, foerster2018counterfactual, ijcai2024p4, li2024agentmixer}. Then during deployment, each autonomous agent only has access to local observations. A primary motivation for the CTDE framework is the nonstationarity of naive decentralized training, as even when a given agent makes no policy changes, another agents' policy modification can vastly alter the team's global reward. The CTDE paradigm's centralized critic can more clearly account for these changes and thus decrease training volatility. Furthermore, the centralized critic can better promote and exploit coordination among agents. Finally, since execution is decentralized, the learned policies for the multi-agent system can also be easily scaled.

A key challenge during the central training phase is how to decouple each policy changes' effects and assess each agent's contribution to the overall success or failure of the global task. Traditionally, the environment provides only a collective reward to the agents based on whether they achieved the shared objective or not. From this single reward, a CTDE training algorithm must determine the contributions of each agent and update the agent policies based on their respective actions~\cite{weiss1995distributed}. This is commonly referred to as the \emph{Structural Credit Assignment} problem~\cite{sutton1984temporal, agogino2004unifying}, not to be mistaken with the temporal credit assignment problem where one must determine the contributions of a succession of actions taken by a single agent receiving a single reward at the end.
Note that the centralized critic must receive and process the joint observations and actions of the agents, which are often of large dimension, and must make assignments based on only partial exploration of the massive joint state-action space, making the contributions difficult to evaluate~\cite{zhou2020learning, yu2022surprising, rashid2020monotonic}. Furthermore, this problem is made more challenging by sparse-reward environments, as the agents must learn to solve an entire sub-task or task before getting any reward feedback. As a result credit assignment has remained an open problem, with other works suffering from limitations such as low-quality feedback, low action influence~\cite{pignatelli2024assessing}, and  difficulty with complex interactions~\cite{lowe2017multi}.

In this work, we reformulate the credit assignment problem as a general pattern recognition problem. More specifically, we view the joint observations and actions of the agents as forming a numerical sequence, which is associated with the global reward signal it earns. Given enough of these sequences, the act of credit assignment is akin to sequence improvement \textemdash evolving beyond the original demonstrations to better achieve goals, generalize to new environments, or increase efficiency/robustness \textemdash in terms of the global reward~\cite{mirchandani2023large}. We propose that this sequential pattern-improvement task can be viewed as giving feedback to each agent regarding their performance during the task. In this work, we claim that the underlying pattern seen in collaborative multi-agent scenarios is akin to an ``Agreement Problem'' where the agents must learn to collectively agree on the strategies needed to effectively  collaborate. With this in mind, we label a few common forms of disagreements seen in multi-agent systems and provide them to our centralized critic. Our primary motivation for posing credit assignment in this pattern recognition form is to leverage human-like skills when assigning credit even in multi-step tasks thanks to their pattern recognition skills and complex reasoning capabilities~\cite{stolyarova2018solving}.

While these abilities have long been a trademark of human intelligence, recent works have shown that foundational Large Language Models~(LLMs) display similar properties~\cite{mirchandani2023large}. Due to the neuromorphic characteristics of this learning architecture, some works argue that LLMs can extract the underlying pattern within the text and act as general pattern machines~\cite{mirchandani2023large} just like humans~\cite{lake2017building}. Further works have shown how LLMs exhibit human-level intelligence and pattern recognition capabilities~\cite{haggstrom2023large, he2024exploring} when evaluated on common human benchmarks such as the International Math Olympiad and the Bar Exam~\cite{perrault2024artificial, barExam, AlphaProof}. Furthermore, LLMs have already been successfully utilized for temporal credit assignment in the single-agent case~\cite{pignatelli2024assessing}. Building on these accomplishments, we propose to leverage the pattern recognition skills of Large Language Models~(LLMs) to tackle the multi-agent structural credit assignment problem.

Specifically, we utilize an LLM as our centralized critic, and instantiate it with textual context of the environment, definitions of our agreement problem, and description of its role. We then periodically provide our critic with joint observations, joint actions, and global rewards from the environment. In return, our LLM-critic provides individualized numerical feedback for each agent and explanations for these assignments. We employ a parser function that takes the feedback from the LLM and maps it to the numerical reward value for each agent at each timestep during the trajectory. We further extend this method by adjusting our query to the LLM as well as our parser function such that during training, each agent can also be given an explicit task from the LLM critic. We call this extension our Large Language Models for Task and Credit Assignment~(LLM-TACA) method. Both our approaches perform exceptionally well, far surpassing current state-of-the-art methods~\cite{zhou2020learning, yu2022surprising, rashid2020monotonic} on common benchmarks~\cite{christianos2020shared, papoudakis2021benchmarking} while also providing more interpretable agent feedback. Lastly, we introduce "Spaceworld," an environment for modeling the multi-agent in-space assembly scenario with safety constraints~\cite{nagpal2023optimal}.

{\flushleft In summary, our main contributions are:}
\begin{enumerate}
    \item A novel reformulation of the credit and task assignment problems as a general pattern recognition problem.
    \item A novel LLM-based centralized training method called ``LLM-MCA'' that performs credit assignment during training and far surpasses the state-of-the-art at many common MARL benchmarks.
    \item An extension of our LLM-critic framework to perform explicit task assignment alongside the credit assignment called ``LLM-TACA'' which further improves performance and explainability.
    \item A new offline dataset with per-robot task and reward annotations to aid future offline-learning efforts, as well as a new MARL benchmark ``Spaceworld'' which simulates a cooperative in-space assembly scenario.
\end{enumerate}

\section{Related Works}
We start by introducing the literature on the credit assignment problem and how large language models have been used in reinforcement learning.

\subsection{Credit Assignment}
One of the main challenges of Multi-Agent Reinforcement Learning~(MARL) is understanding how each agent should be rewarded for their respective participation toward the fulfillment of their shared goal~\cite{weiss1995distributed}. This credit assignment problem arises from the difficulty to discern which agent or action deserves credit when dealing with long time horizons or collaborative tasks~\cite{Sammut2010}. These respective \emph{temporal} and \emph{structural} credit assignment problems have been investigated with reward prediction to assign values to key-actions and adjacent actions~\cite{seo2019rewards}. Some work has also used agent-specific utility functions to demonstrate the equivalence between temporal and structural credit assignment problems~\cite{agogino2004unifying}.

Recent works have built on the notion of difference rewards, which are rewards designed specifically to enable agents to estimate their respective contribution to the shared reward~\cite{nguyen2018credit, feng2022multi, singh2021credit}. Another line of work uses classical cooperative game-theoretic approaches to assign credit to agents~\cite{han2022multiagent, wang2020shapley} while other works have used counterfactual reasoning by comparing the outcomes to what would have happened if agents had not acted~\cite{rahmattalabi2016d++, foerster2018counterfactual}. However, training such a fully centralized state-action value becomes impractical when there are more than a handful of agents.

A more promising subset of work has focused on learning a centralized but factored Q-function, like value decomposition networks~(VDN)~\cite{sunehag2018VDN}, QMIX~\cite{rashid2020monotonic}, and QTRAN~\cite{son2019qtran}, where the agents' local Q-functions can be extracted from a centralized Q-function. However, the structural assumptions enabling this factorization limit the complexity of the centralized action-value functions and often leads to subpar performance~\cite{mahajan2019maven}. This observation led to the development of MAVEN~\cite{mahajan2019maven} to fight inefficient exploration of Q-DPP~\cite{yang2020QDPP}, which bypasses structural factorization methods by encouraging agents to acquire sufficiently diverse behaviors that can be easily distinguished and then factorized. Similar to QMIX, LICA~\cite{zhou2020learning} uses a hyper-network to represent the centralized critic but without the monotony assumption of QMIX~\cite{rashid2020monotonic}. While not explicitly designed to solve the credit assignment problem, multi-agent PPO~(MAPPO)~\cite{yu2022surprising} has shown strong performance in a variety of MARL tasks and will serve as one of our baselines.

\subsection{Large-Language Models in Reinforcement Learning}
Large language models are Transformer-based language models~\cite{vaswani2017attention} trained on immense amounts of text and capable of providing impressive results for a wide array of tasks including text generation~\cite{brown2020language}, question answering~\cite{kenton2019bert}, and text summarization~\cite{zhang2020pegasus}. In the field of reinforcement learning, LLMs have often been used to generate training priors and for improving agent  generalization to new domains~\cite{kim2024survey}. LLMs have also shown promise as zero-shot planners~\cite{huang2022language}, hierarchical planners~\cite{taniguchi2024hierarchical, kannan2023smart}, reward shapers~\cite{carta2022eager}, and reward function generators~\cite{goyal2019using} based on natural language instruction. Closely related to this work, LLMs have also been utilized both as binary critics for solving the temporal credit assignment problem~\cite{pignatelli2024assessing}, and used to help linguistic agents coordinate sub-task plans and waypoint paths through rounds of discussions~\cite{mandi2024roco}.

LLMs have enabled improvements in reinforcement training, including facilitating inter-agent coordination by dictating high-level task planning~\cite{zhuang2024yolo} and negotiation processes~\cite{chen2023multi, sun2024llm}. Additionally, approaches like \cite{slumbers2024leveraging} have used a centralized critic to allow for natural language communication and \cite{hongmetagpt} employs a message pool that agents use to gather relevant messages. While we draw from these works, our approach uses numeric feedback to train neural network policies, as opposed to working with linguistic agents.

\section{Our Method}
While the centralized training aspect of the centralized-training decentralized-execution~(CTDE) paradigm enables agents to learn to coordinate, it also gives rise to the credit assignment problem~\cite{foerster2016learning}. Indeed, distributing the shared reward among agents based on their respective contributions towards the overall objective requires a deep understanding of the shared task. To address this problem, we introduce a novel LLM-based centralized critic to perform credit and task assignments during training.

\subsection{Problem Formulation}
We formalize our scenarios as decentralized partially observable Markov decision processes~(Dec-POMDPs) $\mathcal{M} := \left\langle \mathcal{O},\! \mathcal{S},\! \mathcal{A}, Z,\! T\!, R, \rho \right\rangle$, where $\mathcal{O}$, $\mathcal{S}$, and $\mathcal{A}$ are respectively the joint observation, state, and action spaces, $T: \mathcal{S} \times \mathcal{A} \rightarrow \mathcal{S}$ models the transition probabilities, $Z: \mathcal{S} \rightarrow \mathcal{O}$ is the observation function, $R: \mathcal{O} \times \mathcal{A} \rightarrow \mathbb{R}$ denotes the global reward function, and $\rho$ denote the distribution of initial states $s_0 \in \mathcal{S}$. 
We consider $N$ agents indexed by the set $\mathcal{N} := \{1, \ldots, N\}$. At timestep $k$, agent $i \in \mathcal{N}$ is in a state $s^i_k \in \mathcal{S}^i$ from which it only observes $o^i_k = Z^i\big(s^i_k\big) \in \mathcal{O}^i$ and takes action $a^i_k \sim \pi^i\big(a^i_k | o^i_k\big) \in \mathcal{A}^i$ following its decentralized policy $\pi^i: \mathcal{O}^i \rightarrow \mathcal{A}^i$. The joint observation, state, and action spaces are simply the product of their corresponding subspaces for each agent: $\mathcal{O} := \mathcal{O}^1 \times \mathcal{O}^2 \times \ldots \mathcal{O}^N$, $\mathcal{S} := \mathcal{S}^1 \times \ldots \mathcal{S}^N$, and $\mathcal{A} := \mathcal{A}^1 \times \ldots \mathcal{A}^N$.

At any timestep $k$, the joint action of the agents $a_k := \big(a^0_k, \ldots, a^N_k\big)$ takes their joint state $s_k := \big(s^0_k, \ldots, s^N_k\big)$ to $s_{k+1} \sim T\big(s_{k+1}\, |\, s_k, a_k\big)$. At each time step $k$, the team of agents receives a single collective scalar reward $r_k := R\big(o_k, a_k\big) \in \mathbb{R}$. We denote the joint set of decentralized agent policies by $\pi_\theta := \big(\pi^1, \ldots, \pi^N \big) : \mathcal{O} \rightarrow \mathcal{A}$, where $\theta$ denotes the aggregate of the parameters describing each of the $N$ neural networks modeling the $N$ different policies. 
Our objective is to maximize the expected total reward over this joint policy:
\begin{equation}\label{eq:objective}
    \underset{\theta}{\max} \underset{\substack{s_0\, \sim\, \rho\\a_k\, \sim\, \pi_\theta(o_k)}}{\mathbb{E}} \left( \sum_{k\, =\, 0} R\big(o_k, a_k\big) \right),
\end{equation}
where $o_{k+1} \sim Z\big(T(s_{k+1}\, |\, s_k, a_k)\big)$. The challenge in optimizing~\eqref{eq:objective} resides in coordinating policy updates when changing one set of policy parameters affects the reward earned by the whole team, which leads us to our problem of interest.

\begin{problem}[Credit Assignment]\label{prob:credit_assignment}
    How can we determine the appropriate credit $c_k^i$ to assign to each policy $\pi^i$ at each timestep $k$ given the shared reward $r_k$?
\end{problem}

To address Problem~\ref{prob:credit_assignment} in the CTDE paradigm, a centralized critic distributes the credit $c_k^i$ to each agent $i$. At the beginning of training, this critic first rewards agents for simply learning to operate under the rules of the environment. Once this knowledge is ingrained in the agents, they can begin to solve simple sub-tasks.

At the next stage of training, collaboration should start arising among the agents which additionally introduces intricate issues regarding whether our decentralized agents are collaborating either too little or too much. We say that a joint set of policies $\pi_\theta$ suffers from \emph{under-collaboration} (respectively \emph{over-collaboration}) if a number $j$ of its decentralized agents coordinate to attempt a sub-task requiring $m > j$ (resp. $m < j$) agents. A large challenge in cooperative MARL scenarios is overcoming the initial under-collaboration of decentralized agents. However, an excessive push in this direction will lead to the opposite pitfall of over-collaboration resulting in inefficiencies and agent conflicts. The optimal joint policy $\pi_\theta$ strikes the right amount of collaboration to achieve efficiency when the correct number of agents agree to adopt collaborative strategies. We define this challenge as the ``Agreement Problem''.

It is rather clear that solutions to this ``Agreement Problem'' are the most efficient outcomes, and thus are associated with the strategy producing the largest global cumulative reward $\sum_{k=1} r_k$ from the environment. Our goal, then, is to map a full sequence of observations, actions, and global reward values from our environment to a set of individualized feedback signals, which reflect how well each agent ``agrees'' to the same strategy. In essence, we ask our centralized critic to observe this sequence, propose improvements, and then map those improvements back to the agents that can perform them. Under this lens of the ``Agreement Problem'', we pose the credit assignment problem as a sequence improvement problem, which is then followed by an attribution step, where the critic must choose which among the agents are helping or hurting the overall agreement toward the correct strategy. 

Now that we have re-framed the credit assignment problem as an agreement problem, we will introduce our solution.

\subsection{Large Language Models for Multi-agent Credit Assignment~(LLM-MCA)}\label{sec:Agreement}
Following our reintroduction of Problem~\ref{prob:credit_assignment} into a sequence improvement problem and an attribution step, both of which are pattern recognition tasks, we propose to leverage the strength of Large Language Models~(LLMs) as general pattern machines~\cite{mirchandani2023large}, by designing a novel multi-agent LLM-based centralized-critic architecture, as illustrated in Figure~\ref{fig:diagram}.

\begin{figure}[t!]
    \centering
    \includegraphics[width=\linewidth]{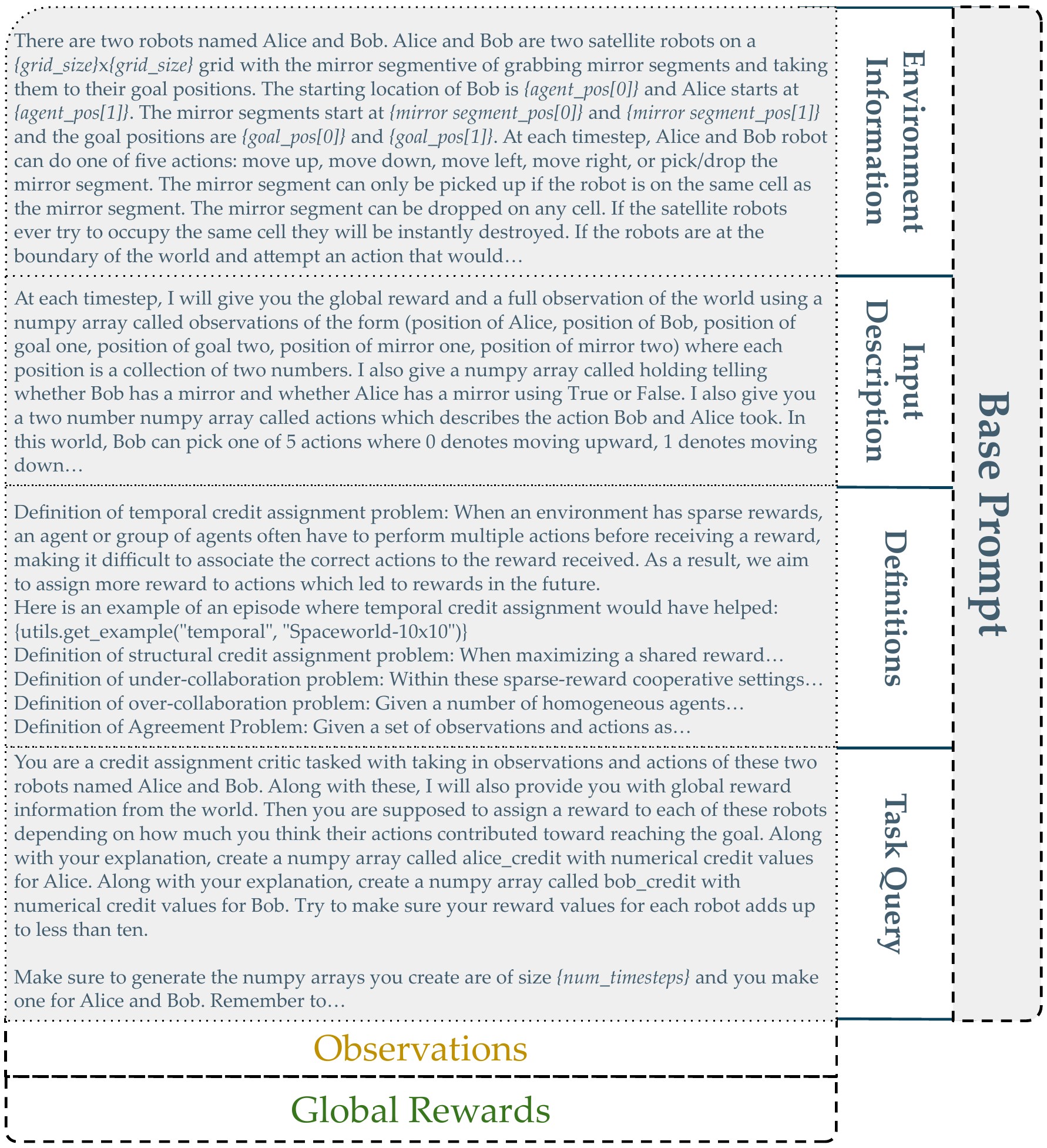}
    \caption{Example prompt for LLM-MCA in the ``Spaceworld'' benchmark. \textmd{Our base prompt $p_\text{base} := (p_\text{env}, p_\text{desc}, p_\text{defn}, p_\text{task})$ is divided into (1) a description of the scenario's rules and objectives, (2) a description of the kinds of inputs it will now receive from that environment, (3) our agreement problem definitions with examples, and (4) a description of its role as a credit assignment agent along with the formatting requirements of its output.}}
    \label{fig:prompt MCA}
    \Description{Sample prompt}
    \vspace{-5mm}
\end{figure}

As previously discussed, at the beginning of training, the primary goal of our LLM credit assignment critic is to adjust the sparse rewards from the environment by implicitly generating sub-tasks and rewarding the agents whenever they achieve these sub-goals. In essence, the LLM critic makes the reward signal denser by providing intermediary rewards. This step is similar to the single-agent temporal credit assignment problem in which LLMs have been shown to work well~\cite{pignatelli2024assessing}. This process inherently requires an understanding of the overall goal of the scenario, which we provide via our initial text prompt. As current methods are unable to effectively utilize this prior, our approach immediately has an advantage. Furthermore, the LLM gets its own feedback via the global reward signal it receives from the environment. To avoid both of the pitfalls of under- and over-collaboration, we inform the LLM critic of the existence of these pitfalls directly within our base prompt as seen in Figure~\ref{fig:prompt MCA}.

These text descriptions of common issues enable the LLM critic to utilize our vocabulary to explain its feedback. We also provide an example trajectory along with these definition prompts to help the LLM understand the issue. Similarly, the agreement problem is explained to the LLM in the form of a concrete example in its current environment, as reproduced below.
\begin{prompt}\label{prompt:agreement}
    Given a set of observations and actions as performed by the agents in the environment, there will be times when the agents will individually accomplish goals, but occasionally, we will need them to collaborate. When two agents correctly agree to collaborate on a task that requires two agents, then they have found a valid solution to the ``Agreement Problem''.
\end{prompt}

Specifically, we instantiate the LLM critic with a base prompt $p_\text{base} \in \mathcal{P}$ where $\mathcal{P}$ is the prompt space. This base prompt is a concatenation of four sub-prompts $p_\text{base} := (p_\text{env}, p_\text{desc}, p_\text{defn}, p_\text{task})$ tailored for the given environment. The first element $p_\text{env}$ is a description of this scenario and the rules of the environment. This also includes a description of the scenario's objective and what will garner the agents better rewards. $p_\text{desc}$ is an explanation of the kinds of data it will receive from this environment and the agent policies. Following our discussion above, we also provide the LLM critic with a Definition Prompt for under-collaboration, over-collaboration, and the Agreement problem~\ref{prompt:agreement}. Combining these with traditional definitions of the temporal credit assignment and structural multi-agent credit assignment problems, we have our definition prompt $p_\text{defn}$. We also provide the LLM with a task description $p_\text{task}$ which details its role and objective in the training process, as well as the precise form of its expected output at each timestep. Then, at each training timestep, we provide our LLM-critic $C_{LLM}$ with dynamic data from the environment consisting of past observations, actions, and rewards. See Figure~\ref{fig:prompt MCA} for a more detailed example of our prompt for the ``Spaceworld'' benchmark.

At each timestep of the training process, we provide the next observation $o_{k+1} = Z(f(s_k, a_k)) \in \mathcal{O}$ and global reward $R(o_k, a_k) \in \mathbb{R}$ in the form of a prompt $p_{k}$. Due to the nature of LLM conversations, each new prompt $p_{k+1}$ is effectively augmented with the history of past prompts and feedback, along with the new data, such that $p_{k+1} := \big(p_k, R(o_{k}, a_{k}), o_{k+1}, a_{k+1}\big)$ where $p_0 := \big( p_\text{base}, o_0, a_0\big)$. The LLM-critic then provides a numerical reward value $c^i_k$ to each agent in the scenario, along with an explanation of how it generated the feedback for that agent. This explanation is often associated with our ``Agreement Problem'' paradigm and how the LLM's feedback helps mitigate disagreements among agents. An example of this behavior can be seen in Figure~\ref{fig:TACA-output}

As our LLM-critic $C_{LLM}: \mathcal{P} \times \mathbb{R} \times \mathcal{O} \times \mathcal{A} \rightarrow \mathcal{P}$ maps only to the text space, we utilize a parser function $F_{MCA}$ which takes the entire output of the LLM critic and extracts the array of individualized credit values $c_k \in \mathbb{R}^N$. Because we include requirements for the format of this array within our base's task prompt $p_\text{task}$, our parser function is comprised of a simple regex expressions search and a normalization step. As such, we denote our parsed LLM-critic $F_{MCA} \circ C_{LLM}: \mathcal{P} \times \mathbb{R} \times \mathcal{O} \times \mathcal{A} \rightarrow \mathbb{R}^N$, and explicitly have $F_{MCA}\big(C_{LLM}\big(p_k, r_k, o_{k+1}, a_k\big)\big) = \big(c_{k}^1, \ldots, c_{k}^N\big)$, where each $c_{k}^i$ is the credit assigned to each agent $i \in N$ for their contribution to the shared reward $r_k$.

Then, instead of directly attempting to maximize~\eqref{eq:objective}, we optimize the surrogate objective generated by our LLM critic:

\begin{equation}\label{eq:surrogate_objective}
    \underset{\theta}{\max} \underset{\substack{s_0\, \sim\, \rho\\a_k\, \sim\, \pi_\theta(o_k)}}{\mathbb{E}} \left( \sum_{k\, =\, 0} \sum_{i\, =\, 1}^N F_{MCA}\big(C_{LLM}\big(p_k, R(o_k, a_k), o_{k+1}, a_k\big)\big)  \right),
\end{equation}
where $o_{k+1} \sim Z\big(T(s_{k+1}\, |\, s_k, a_k)\big)$. This formulation allows our centralized LLM critic to directly evaluate the complex interplay behavior between agents.

Since we optimize our neural-network policy parameters through batch updates, as opposed to after every timestep, we can instead compute our credit assignment only at these policy update steps. As such, our surrogate objective can be further relaxed, where we compute feedback on an entire trajectory or batch of trajectories at once. We illustrate this batch-training process in Fig.~\ref{fig:process}. With this in mind, we adjust our base prompt $p_{\text{base}}$ to describe this updated task, concatenate all of the observations, actions, and global rewards to a single prompt $p_{\text{batch}} = (p_{\text{base}}, p_0, p_1, \ldots)$, and extract a set of credit matrices from the LLM critic at once. By providing an entire set of agent trajectories at once, our LLM-critic can better analyze the agents' actions in retrospect, which further enables it to notice more intricate patterns within the agents' behavior. Another major benefit to our batching technique is that it greatly reduces compute times and the data storage burden.

This LLM-MCA method also has several other inherent benefits. Since our LLM-critic is innately language-based, it is also able to explain and justify their feedback strategy, as seen in Figure~\ref{fig:TACA-output}. This explainability contrasts all previous multi-agent credit assignment works, where the crediting strategy often results from an uninterpretable deep neural network optimization, making the process a complete black-box for the user. Furthermore, since our critics are pre-trained foundation models, they do not require any major hyperparameter tuning.

In summary, our LLM-MCA method can directly map entire trajectories of agents to individualized feedback for each agent, in a computationally efficient and interpretable manner. After extracting the credit values from this feedback, we can directly train our set of decentralized policies.

\subsection{Large Language Models for Task and Credit Assignment~(LLM-TACA)}
Based both upon our interpretation of credit assignment as an ``Agreement Problem'' and our analysis of the outputs of our LLM-MCA method, we noticed that during the credit assignment process, our critic often implies a task assignment, and disincentivizes the agent from deviating from its internal task allocation strategy. From the point of view of the agent, whenever their learned task assignment differs from the critic's the credit earned decreases dramatically, causing training volatility. Based on these findings, we introduce our extension LLM-TACA which includes a novel explicit task assignment aspect alongside the centralized credit critic.

\begin{figure}[h!]
    \centering
    \includegraphics[width=0.9\linewidth]{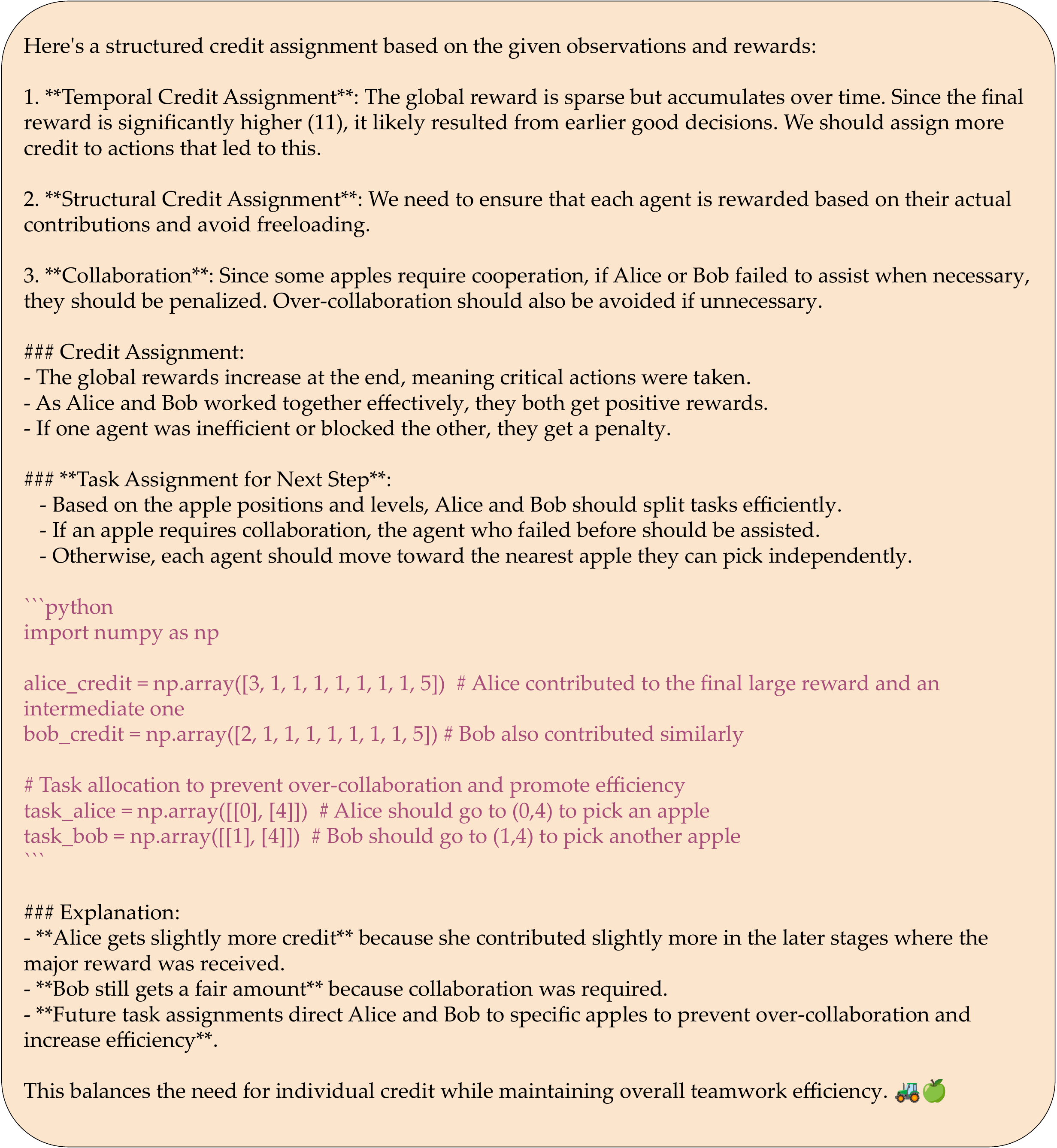}
    \caption{Example output from our LLM-TACA method in the level-based foraging benchmark. \textmd{Our LLM-critic provides individualized credit assignments for the previous timesteps, task assignments, and explanations for its decisions.}}
    \label{fig:TACA-output}
    \Description{Example output from our LLM-TACA method in the level-based foraging benchmark}
\end{figure}

We instantiate our LLM-TACA with an adjusted base prompt $p_\text{base}^T = (p_\text{env}, p^T_\text{desc}, p_\text{defn}, p^T_\text{task})$, where $p^T_\text{desc}$ now also includes a description of the task assignment the LLM is allowed to provide and $p^T_\text{task}$ includes the additional request to generate these individualized task goals. A Practically, we allow LLM-TACA to provide additional inputs directly to the agent policies during training time. An example of this assignment is shown in Figure~\ref{fig:TACA-output}. The exact form of the task inputs is a design choice, but we primarily used short integer arrays. At each timestep, on top of the credit feedback, LLM-TACA was given the option to either provide an additional task assignment input $t_k$ directly to the policy of a given agent, or to provide nothing. To extract these features from the overall feedback and explanation from our LLM, we introduce a new parser function
\begin{equation}
    F_{TACA}(C_{LLM})(p^T_k, R(o_k, a_k), o_{k+1}, a_k) = (c_{k}^1, t_{k}^1, \ldots, c_{k}^N, t_{k}^N).
\end{equation}

However, as this task assignment input is only provided during the centralized training and not during the decentralized execution, we must ensure that our agent policies don't put too much weight on these inputs. To achieve this, we construct our neural network-based policies with a zero-padded input space and controlled dropout, as seen in early neural network literature~\cite{gimenez2020semantic, srivastava2014dropout}. More specifically, over the course of the training, we input fewer and fewer task assignments from LLM-TACA by instead overwriting this data with zeros and also increasing the dropout rate of the neurons associated with these inputs.

In summary, our LLM-TACA method allows us to provide credit assignment and explicit task assignment information directly to each agent's policy during training, while still maintaining the ability to execute them in a decentralized manner.

\section{Results}
Now that we have introduced our two methods LLM-MCA and LLM-TACA for multi-agent credit assignment, we will evaluate them on a range of discrete, sparse-reward environments against state-of-the-art baselines in MARL. In these results, we seek to show the efficacy of our framework rather than that of any specific language foundation model. Thus, we do not compare the performance of different large language models and only use the Gemma-7B model~\cite{team2024gemma} for our trials. In this work, we use an open model to exploit the monetary and ease-of-access benefits, but rudimentary testing suggested that most popular foundation models could be used to generate similar results. Additionally, as all LLMs' reasoning capabilities improve, the performance of our framework should similarly improve. All of the following results use the double DQN architecture~\cite{van2016deep} for the agent's policies, which are trained and run on a NVIDIA 3080 Ti.

Since LLMs generate text, their outputs are interpretable for human users and their choices are thus explainable. This added benefit enables our centralized LLM-critic to explain its decisions when providing reward feedback values, especially how they are aligned with our ``Agreement Problem'' as discussed in Section~\ref{sec:Agreement}. More specifically, with each alteration our LLM-critic makes, it often provides explanations for each of its actions in terms of how they help mitigate one of the problems as defined in the $p_\text{defn}$ part of our base prompt, as illustrated in Figure~\ref{fig:TACA-output}. Furthermore, during early points in training, our LLM critics often rewards the agents for accomplishing some internal sub-goal and mention that they are solving the sparse reward or temporal credit assignment problem.


\begin{figure*}[ht!]
    \centering
    \includegraphics[width=0.9\linewidth]{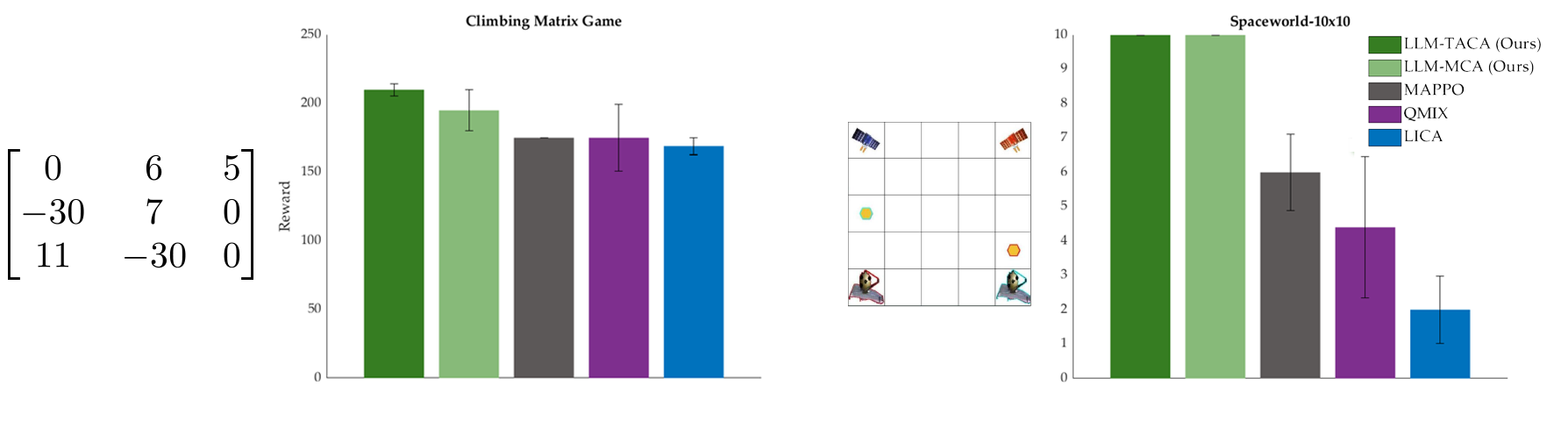}
    \caption{Comparison of our methods with baselines on benchmarks. \textmd{All values reported are averaged across five separate trainings and the 95\% confidence interval is illustrated by the error bars. Spaceworld is our custom benchmark where agents must avoid collision while transporting respective parts to their destinations.".}}
    \label{fig:results}
    \Description{Comparison of our methods with baselines on common benchmarks.}
\end{figure*}


We first test our methods on a fully observable, cooperative climbing matrix game with two agents and dense rewards~\cite{claus1998dynamics}. This repeated matrix game requires both agents to find an optimal policy that maximizes their rewards as described by the game's common-payoff matrix illustrated in Figure~\ref{fig:results}. Each episode is 25 timesteps in length and agents are given constant observations. The objective is for agents to learn to maximize their common-payoff by consistently agreeing to play the action profile which provides the highest common reward. It is easy to see that there exists multiple suboptimal equilibria, and so myopic critics may fail to train for the truly optimal solution~\cite{claus1998matrixgames}.

\begin{figure*}[ht!]
    \centering
    \includegraphics[width=0.9\linewidth]{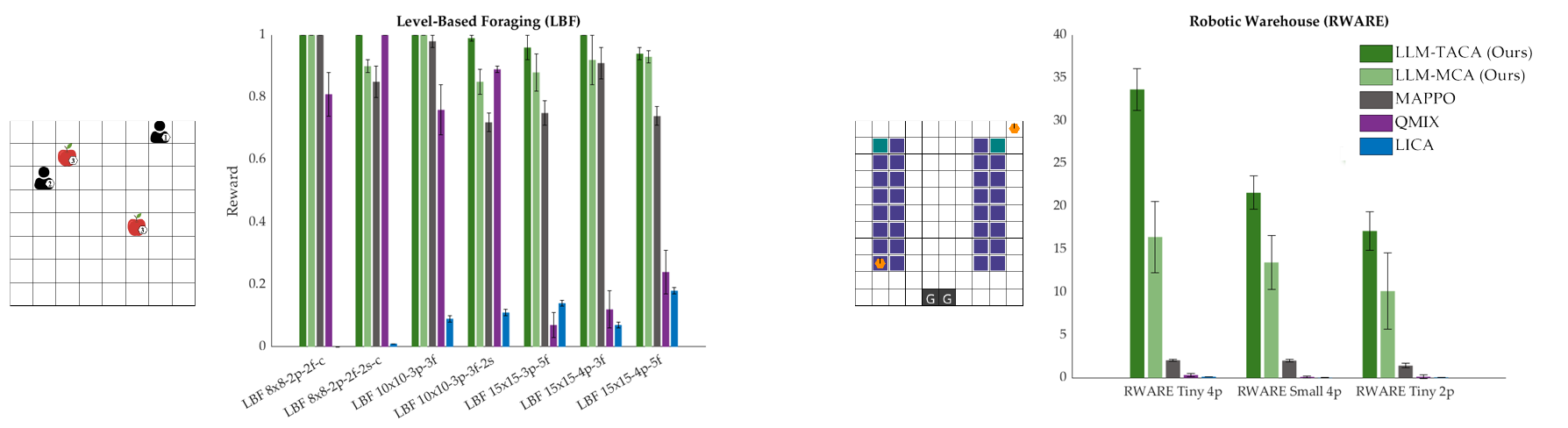}
    \caption{Comparison of our methods with baselines on common MARL benchmarks. \textmd{All values reported are averaged across five separate trainings and the 95\% confidence interval is illustrated by the error bars. We compared our methods on 12 scenarios across two common MARL benchmarks. Level-Based Foraging~(LBF) is a common benchmark where agents must coordinate to pick apples based on their levels and the level of the apple. Robotic Warehouse~(RWARE) is a partially-observable benchmark where robots must move certain packages to the delivery zone. LBF scenario names follow the format: "\{GRID SIZE\}-\{PLAYER COUNT\}p-\{FOOD COUNT\}f-\{SIGHT COUNT\}\{-c IF COOPERATIVE MODE\}". RWARE environment names follow the format "\{GRID SIZE\} \{PLAYER COUNT\}p".}}
    \label{fig:results-bot}
    \Description{Comparison of our methods with baselines on common benchmarks.}
\end{figure*}

Next, we evaluate our approach on the Level-Based Foraging~(LBF) environment~\cite{christianos2020shared, papoudakis2021benchmarking} as seen in Figure~\ref{fig:results-bot}(a). In this partially observable grid-world environment, multiple agents cooperate to collect apples from trees scattered within the area. Each agent has an associated level and to harvest a given apple the combined levels of the agents next to the tree must be greater than or equal to the level of the food. The environment name lets you customize many elements including the number of agents, the observability, the grid size, cooperativity, and the maximum number of food elements. The exact environment parameters we used for our scenarios can be seen directly in the environment names labeled in Figure~\ref{fig:results-bot}.

Our third testing environment is Robotic Warehouse~(RWARE)~\cite{papoudakis2021benchmarking, christianos2020shared} as shown in Figure~\ref{fig:results-bot}(b). In this partially observable environment, robots are tasked with locating and delivering shelves to human workstations before returning them to empty shelf locations. However, as each robot only gets a 3x3 observation centered on themselves it is not always clear which area the agent should traverse. Additionally, there are sparse rewards and agents can get in each others way, causing needless slowdowns and inefficiencies. This environment is available with 3 different overall grid sizes and the number of robots can be specified, both of which are mentioned in the environment labeled in Figure~\ref{fig:results-bot}.

Finally, we introduce a new MARL benchmark called ``Spaceworld'' which simulates a cooperative In-Space Servicing, Assembly, and Manufacturing~(ISAM) scenario~\cite{nagpal2023optimal}. In this grid-world environment illustrated in Figure~\ref{fig:results}(c), the agents are satellite robots tasked with servicing the James Webb Space Telescope by replacing damaged mirrors. The blue and red agents are tasked with bringing the correct mirror segment to their target, as illustrated via coloring. However, both agents are allowed to interact and move the mirror segments wherever they like. The primary motivation behind this benchmark was to include safety consequences and unforced collaboration. If the two agents collide (i.e. try to enter the same square) at any point, both agents are destroyed and the episode is terminated with reward zero. This indirectly incentivizes one agent to stay still and avoid collision while the other agent moves both objects to their required locations. However, this produces suboptimal results as one point is deducted from their global reward for every additional timestep the agents take over the minimal. This reward signal is automatically normalized across grid sizes such that the highest achievable reward is always $10$ and the lowest is $0$. As such, the centralized critic must balance the safety and efficiency objectives to gain the highest possible reward.

We believe that these benchmarks effectively showcase our methods' capabilities for enabling intricate interactions in sparse-reward environments, handling partial observability, and learning safety-critical behaviors.

\subsection{Baselines}
To highlight the performance of both of our methods, LLM-MCA and LLM-TACA, we compare them against a number of other Centralized-Training Decentralized-Execution approaches.

First, we compare with LICA~(Learning Implicit Credit Assignment for Cooperative Multi-Agent Reinforcement Learning)~\cite{zhou2020learning}, a multi-agent actor-critic method representing the centralized critic with a hyper-network. This approach is motivated by finding locally optimal cooperative behaviors by directly maximizing a joint action value function.

Our second baseline is MAPPO~(Multi-Agent PPO), an extension of PPO to multi-agent scenarios with centralized value function inputs~\cite{yu2022surprising}. This multi-agent actor-critic method learns the joint state value function and operates in cooperative, common-reward settings.

Finally, QMIX~\cite{rashid2020monotonic} is an extension of value decomposition networks~(VDN)~\cite{sunehag2018VDN} to more complex centralized action-value functions. It uses a mixing network to estimate joint action values as a nonlinear combination of individual agent values.

\subsection{Results Analysis}
We compiled the results of our method against our 3 baselines in Figures~\ref{fig:results} and \ref{fig:results-bot}. LICA completely fails to solve the matrix game, while all other methods achieve high scores dominated by our LLM-MCA and LLM-TACA methods. Already with the matrix benchmark, we see the benefit of having a critic which easily incorporates prior environment knowledge and retrospective analysis.

On the Spaceworld environment with a $10 \times 10$ grid both our methods consistently achieve the highest score whereas the three baselines have significantly lower scores and higher variance over 5 runs as they either succeed or completely fail. Due to any collisions immediately ending an episode, agents must learn to avoid such scenarios consistently and in a sample efficient manner. Additionally, as starting, object, and goal locations are randomized between episodes, our baselines often faced issues with nonstationary policy changes causing poor performance in even previously seen scenarios. 

On the Level-Based Foraging environment both our methods consistently score in the 90th percentile with low variance with an advantage for LLM-TACA. MAPPO and QMIX achieve high scores on the smallest grid sizes but their performance drops, drastically for QMIX, as the grid size increases. On the other hand, LICA performs consistently poorly on all sizes.

In the Robotic Warehouse, the results exhibit a similar trend with both our approaches achieving scores an order of magnitude higher than that of MAPPO, which scores an order of magnitude higher than QMIX and LICA.

On the whole, as the tasks become more complex, the impact and efficiency of our method become more evident. Our method was able to harness prior knowledge about the scenarios to better assess agent performance and provide dense feedback quickly, causing our learned policies to be far less myopic than those produced by the baselines.

\subsection{Offline Dataset}
As a byproduct of our LLM critic method, we have gathered a large number of trajectories in each of these scenarios along with the individualized credit and task assignment for each agent during each timestep of the training process. This novel offline dataset will enable decentralized offline-learning future efforts\footnote{Dataset instructions at http://iconlab.negarmehr.com/LLM-MCA/}. While there have been efforts to learn multi-agent policies from offline data, they often must rely on an underlying centralized structure~\cite{jiang2023offline, wang2024offline}, and so we provide this dataset with the hope that further progress can be made in this direction.

\section{Conclusions and Future Work}
In this work, we proposed two LLM centralized critics for solving the multi-agent credit assignment problem. Specifically, we reformulated this credit assignment problem as a general sequence improvement problem. Then, building on humans' innate skills at solving general pattern recognition problems, we noticed a similar skill emerging in the latest LLMs. We leveraged this skill by introducing two LLM-based centralized critics tasked with assigning credit to each agent based on their respective participation toward their shared objective.

We evaluated our approach on major MARL benchmarks where we surpassed the state-of-the-art by a significant margin. Thanks to the interpretability and explainability of LLMs, the users of our method can easily understand both the emergent behaviors observed by the LLM critic among the agents and the feedback the LLM is proposing to improve their capabilities.

A limiting factor to this work is the LLM's slowness in generating outputs. We have already worked on mitigating this issue via batch processing of trajectories to minimize the number of LLM queries. Another limitation is the monetary cost of using closed-source LLMs like ChatGPT.

We envision several avenues for future work. We first plan to expand our approach to non-cooperative tasks as they can also suffer from the credit assignment problem. We want to utilize faster LLMs to enable real-time evaluations. This advance would then facilitate the deployment of our method on real robots.

\begin{acks}
This material is based upon work supported by the National Science Foundation under grants DGE-2146752, CNS-2423130, CCF-2423131, and ECCS-2145134 CAREER Award. 
\end{acks}

\newpage
\bibliographystyle{ACM-Reference-Format} 
\bibliography{references}

\newpage
\section{Appendix}

\subsection{Implementation Details}

Note that when using the Level-Based Foraging~(LBF) environment\footnote{Note that the Level-Based Foraging environment has since changed the "-c" tag to "-coop" in version 3.} \cite{christianos2020shared, papoudakis2021benchmarking} the environment name lets you customize many elements including the number of agents, the observability, the grid size, cooperativity, and the maximum number of food elements. These elements are defined in the environment name in \ref{table:results} with format: "\{GRID SIZE\}-\{PLAYER COUNT\}p-\{FOOD COUNT\}f-\{SIGHT COUNT\}\{-c IF COOPERATIVE MODE\}".

Additionally, for our Robotic Warehouse~\cite{papoudakis2021benchmarking, christianos2020shared} scenarios, this environment is available with 3 different overall grid sizes and the number of robots can be specified, both of which are mentioned in the environment name with the format "\{GRID SIZE\} \{PLAYER COUNT\}p".

We summarize our main results in Table~\ref{table:results}.

\begin{table*}[ht!]
    \caption{Comparison of our method with previous solutions on common benchmarks. All values reported are the means and 95\% confidence interval for the collective return of the agents by each algorithm over 5 separate training runs with no common reward or reward scalarization techniques. The \textbf{bold} score denotes the highest value of a row and an asterisk\nb denotes the second highest.
    }
    \label{table:results}
    
    \resizebox{\textwidth}{!}{%
    \begin{tabular}{cccccc}
        \toprule Environments\textbackslash Algorithms & LLM-MCA (Ours) & LLM-TACA (Ours) & LICA~\cite{zhou2020learning} & MAPPO~\cite{yu2022surprising} & QMIX~\cite{rashid2020monotonic} \\
        \midrule
        Climbing Matrix Game & $195.0 \pm 33.5$\nb & $\mathbf{210.0 \pm 10.0}$ & $169.0 \pm 13.4$ & $175.0 \pm 0.0$ & $175.0 \pm 54.8$ \\
        Spaceworld-10x10 & $\mathbf{10.0 \pm 0.0}$ & $\mathbf{10.0 \pm 0.0}$ & $2.0 \pm 2.2$ & $6.0 \pm 2.5$\nb & $4.4 \pm 4.6$ \\
        \hdashline
        Level Based Foraging & & & & & \\
        8x8-2p-2f-c     & $\mathbf{1.00 \pm 0.00}$  & $\mathbf{1.00 \pm 0.00}$ & $0.00 \pm 0.00$ & $\mathbf{1.00 \pm 0.00}$ & $0.81 \pm 0.07$\nb \\
        8x8-2p-2f-2s-c  & $0.90 \pm 0.02$\nb        & $\mathbf{1.00 \pm 0.00}$ & $0.01 \pm 0.00$ & $0.85 \pm 0.05$           & $\mathbf{1.00 \pm 0.00}$\\
        10x10-3p-3f     & $\mathbf{1.00 \pm 0.00}$  & $\mathbf{1.00 \pm 0.00}$ & $0.09 \pm 0.01$ & $0.98 \pm 0.02$\nb        & $0.76 \pm 0.08$\\
        10x10-3p-3f-2s  & $0.85 \pm 0.04$           & $\mathbf{0.99 \pm 0.01}$ & $0.11 \pm 0.01$ & $0.72 \pm 0.03$           & $0.89 \pm 0.01$\nb \\
        15x15-3p-5f     & $0.88 \pm 0.06$\nb        & $\mathbf{0.96 \pm 0.04}$ & $0.14 \pm 0.01$ & $0.75 \pm 0.04$           & $0.07 \pm 0.04$\\
        15x15-4p-3f     & $0.92 \pm 0.08$\nb        & $\mathbf{1.00 \pm 0.00}$ & $0.07 \pm 0.01$ & $0.91 \pm 0.05$           & $0.12 \pm 0.06$\\
        15x15-4p-5f     & $0.93 \pm 0.02$\nb        & $\mathbf{0.94 \pm 0.02}$ & $0.18 \pm 0.01$ & $0.74 \pm 0.03$           & $0.24 \pm 0.07$\\
        Average         & $0.93 \pm 0.03$\nb        & $\mathbf{0.98 \pm 0.01}$ & $0.09 \pm 0.01$ & $0.85 \pm 0.07$           & $0.54 \pm 0.05$ \\
        \hdashline
        Robotic Warehouse & & & & & \\
        Tiny 4p     & $16.44 \pm 4.16$\nb  & $\mathbf{33.65 \pm 2.44}$ & $0.14 \pm 0.02$ & $2.05 \pm 0.12$ & $0.33 \pm 0.18$ \\
        Small 4p    & $13.47 \pm 3.15$\nb  & $\mathbf{21.59 \pm 1.95}$ & $0.03 \pm 0.01$ & $2.01 \pm 0.18$ & $0.12 \pm 0.11$ \\
        Tiny 2p     & $10.12 \pm 4.43$\nb  & $\mathbf{17.13 \pm 2.27}$ & $0.06 \pm 0.02$ & $1.43 \pm 0.26$ & $0.14 \pm 0.19$ \\
        Average     & $13.34 \pm 3.91$\nb  & $\mathbf{24.12 \pm 2.22}$ & $0.08 \pm 0.02$  & $1.83 \pm 0.19$ & $0.20 \pm 0.16$ \\
        \bottomrule
    \end{tabular}}
\end{table*}

\end{document}